\documentclass[sigconf]{acmart}
\usepackage{tabularx}
\usepackage{multirow}

\newcolumntype{L}{>{\raggedright\arraybackslash}X}

\usepackage[colorinlistoftodos]{todonotes}
\AtBeginDocument{%
  \providecommand\BibTeX{{%
    \normalfont B\kern-0.5em{\scshape i\kern-0.25em b}\kern-0.8em\TeX}}}

\begin{document}

\title{Know Your Model (KYM): Increasing Trust in AI and Machine Learning}

\author{Mary Roszel}
\affiliation{%
  \institution{University of Luxembourg}
  \streetaddress{29, Avenue J.F. Kennedy, L-1855 Luxembourg}
  \country{Luxembourg}
  \postcode{L-1855}
}
\email{mary.roszel@uni.lu}

\author{Robert Norvill}
\affiliation{%
  \institution{University of Luxembourg}
  \streetaddress{29, Avenue J.F. Kennedy, L-1855 Luxembourg}
  \country{Luxembourg}
  \postcode{L-1855}
}
\email{robert.norvill@uni.lu}

\author{Jean Hilger}
\affiliation{%
  \institution{Banque et Caisse d’Epargne de l’Etat}
  \streetaddress{Luxembourg}
  \country{Luxembourg}
}
\email{jean.hilger@ext.uni.lu}

\author{Radu State}
\affiliation{%
  \institution{University of Luxembourg}
  \streetaddress{29, Avenue J.F. Kennedy, L-1855 Luxembourg}
  \country{Luxembourg}
  \postcode{L-1855}
}
\email{radu.state@uni.lu}






\begin{abstract}
The widespread utilization of AI systems has drawn attention to the potential impacts of such systems on society. Of particular concern are the consequences that prediction errors may have on real-world scenarios, and the trust humanity places in AI systems. It is necessary to understand how we can evaluate trustworthiness in AI and how individuals and entities alike can develop trustworthy AI systems. In this paper, we analyze each element of trustworthiness and provide a set of 20 guidelines that can be leveraged to ensure optimal AI functionality while taking into account the greater ethical, technical, and practical impacts to humanity. Moreover, the guidelines help ensure that trustworthiness is provable and can be demonstrated, they are implementation agnostic, and they can be applied to any AI system in any sector.
\end{abstract}

\maketitle

\section{Introduction}\label{sec:intro}
Artificial intelligence (AI) systems are being utilized in nearly every sector, with notable applications in autonomous systems, banking, education, medical, manufacturing, and robotics. 
In recent years, advancements in the field of AI have drawn attention to concerns of the large-scale impacts of such AI systems \citep{amodei2016concrete}, urging awareness of the potential harm that these systems may cause. With the increasing utilization of AI systems, particularly in high-stakes areas such as autonomous vehicles, health-care services, and surveillance, it is vital that we consider their trustworthiness.  

Consider, for example, the use of AI in medical diagnoses and decisions where an AI assists medical personnel in making decisions based on diagnostic scans. In theory, these systems would be of great use by allowing for faster and more accurate diagnoses, allowing hospitals to treat additional patients, and reduce human error. However, while these systems perform well in controlled environments, in real-world applications they often perform poorly and increase the time required to make the diagnosis \citep{beede2020human}. In this example, the issue of trust in such a system comes into consideration on multiple levels: the system must have the trust of the medical personnel utilizing it, the patient, and its greater community. How do we trust that the decisions these systems make are accurate, and who takes the blame if a patient is misdiagnosed, or if time is wasted on attempting to use a faulty system?  

Trustworthiness of an AI system implies that the development of such a system considers the greater ethical, technical, and practical impacts to humanity. Our paper frames trustworthy AI around eight key principles as discussed in \citet{fjeld2020principled}: accountability, fairness and non-discrimination, human control of technology, privacy, professional responsibility, promotion of human values, safety and security, and transparency and explainability. To encourage large-scale AI adoption and increase trust, the burden is on the creators should address these principles in their deployed systems. However, developing a completely trustworthy AI system is a difficult task. Currently, there is no formal method for tracking and reporting how developers address issues of trust. 
Without such a method, the wide-scale development of trustworthy AI systems is greatly hindered.  

Further, the development of such systems may lead to problems including scalability and compliance; 
with the development of complex models, how do companies and individuals ensure their systems are trustworthy, and at what cost?
For example, in the financial sector, Know Your Customer (KYC) refers to the legal requirements institutions have to ensure they know the true identity of their clients. State-of-the-art technologies have been successfully leveraged to ease the burden of compliance~\cite{norvill2020security}. We envisage a similar regulatory landscape for the use of AI tools whereby institutions utilizing AI must be able to prove their tools can be trusted. Moreover, ensuring compliance with current and future legislation will enable the data economy to operate, ensuring AI is put to the best possible use. 

Toward developing trust in AI (and easing regulatory burden), we propose the concept of Know Your Model (KYM), the idea that all models have a unique identity and that model characteristics can be leveraged to know and trust models. The concept of KYM is influenced by the idea that all models have a unique identity and that model characteristics can be leveraged to know and trust models. To "know" a model implies collecting, recording, and storing detailed records of the processes undergone during the development of a model, subsequently establishing \textit{model identity}.

To know a model, we propose 20 key guidelines that creators should address to establish a model's identity, particularly in respect 4 core concepts: \textbf{efficacy}, \textbf{reliability}, \textbf{safety}, and \textbf{responsibility}. These guidelines provide a general framework that is applicable to any and all AI implementations, rather than prescribing a particular implementation. The proposed guidelines are concise suggestions of important aspects that creators should be able to address about their AI systems in regards to processes, methodology, and trust. These guidelines can be leveraged by creators to increase transparency and trustworthiness in their AI development processes.

Our aim is not to provide a definitive solution for developing trust in AI, rather 
to suggest key areas for increased attention in development, considering technical, ethical, and legal aspects in addition to trust. Therefore, the primary aim of this paper is to outline a method to establish model identity with a general framework that all creators can apply to AI system development. 
The information required to fulfill the guidelines will vary by the complexity of each system, with more complex systems requiring greater attention to nuances in their use of data and modeling processes. This attention to detail will benefit creators by ensuring that the appropriate information is collected during each stage of AI development and easing the burden of proof for the effectiveness and trustworthiness of their systems.

The rest of this paper is structured as follows. Section \ref{sec:trust} discusses the current field of trustworthy AI and eight principles frequently observed in trustworthy AI research. Section \ref{sec:related} presents related work in tracking the development of AI systems. Section \ref{sec:KYM} provides an overview of KYM, emphasizing the need for efficacy, reliability, safety, and responsibility. This section describes in detail the 4 core concepts of KYM.  Section \ref{sec:requirements} lists the guidelines KYM and how developers can leverage them. Finally, Section \ref{sec:discussion} summarizes the contributions in this paper and discusses the challenges and future work needed to establish KYM and increase trust in AI.







\section{Trustworthy AI} \label{sec:trust}
The domain of trustworthy AI has gained traction in recent years with an increase in concern about the impact of AI deployment on society. When developing AI systems we often consider its accuracy in decision making, but accuracy alone is not enough in high-stake scenarios (such as judicial decisions, and fraud detection) where an incorrect decision may have undesirable consequences. 
The large-scale adoption of AI systems greatly depends on developing trust in not only their performance but also their greater purpose and transparency. Developing trust in AI is a dynamic process, it is crucial to continuously develop and maintain trust throughout all stages of development and deployment \citep{siau2018building}.


 
 
While no consensus has been found on the formal definition of trustworthy AI, focus has been placed on eight key principles frequently observed in the domain: accountability, fairness and non-discrimination, human control of technology, privacy, professional responsibility, promotion of human values, safety and security, and transparency and explainability \citep{fjeld2020principled}. Increasing trust in AI requires that companies and developers closely analyze how they address these key principles during the development of their systems. In this section, we outline these key areas and the challenges they pose in the development of trustworthy AI systems. 

\subsection{Accountability}
With AI becoming increasingly prevalent in society, there is increasing concern about who will be accountable for the decisions and impact of AI technologies. The principle of accountability calls for the verifiability and replicability of AI systems, as well as calling attention to the need for auditing, regulatory, and legal requirements and responsibility \citep{fjeld2020principled}. 

In particular, the verifiability and replicability of AI systems are paramount for building trust. Developers should ensure that all experiments and results are replicable (reproducible), and provide enough information to be able to be validated. Without this information, users may find the systems, and organizations who made them, untrustworthy. 

Further, the principle of accountability is particularly important when considering regulatory conditions. For regulatory bodies standards for accountability in AI remain an open question as they seek a balance between creator responsibility and taking full advantage of the capabilities of AI \citep{doshi2017accountability}. 

\subsection{Fairness \& Non-Discrimination}
The fairness and non-discrimination principle calls for the consideration, detection, and prevention of discrimination and bias in the development of AI systems. As these systems are often used in sensitive or high-stakes areas, it is vital that decisions are made without discriminatory or biased influence toward particular demographic groups or populations. Biased and discriminatory practices in AI have been identified in nearly every type of system, including, advertisement, chatbots, employment decisions, legal decisions, facial and voice recognition, and search engines \citep{mehrabi2019survey}. Creators need to consider how their systems make decisions and potential harmful effects they may cause.


Research in this area provides many solutions for auditing, and improving bias and fairness in AI systems. Typically, bias is unintentionally introduced with biased training data and can be addressed with statistical methods applied to data or models \citep{bellamy2018ai}. 
For a complete survey of such methods, we suggest the article by \citet{mehrabi2019survey}. 





\subsection{Human Control of Technology}
With the increasing amount of AI systems making sensitive and high-stakes decisions, it is vital to consider where we shift control of decision-making from humans to AI. Consider the case of autonomous vehicles: self-driving vehicles are able to make decisions and operate without human control, but there are continual ethical concerns about the decisions made in accident-scenarios \citep{nyholm2016ethics}.

Research in this area calls for the ability for humans to review the decisions made by AI, or for AI to be built with the ability for humans to intervene, especially in the case of dangerous or costly decisions \citep{hook2000steps}. In particular, the field of human-computer interaction includes a great body of work on human control in AI but has found it difficult to establish guidelines for the design of AI systems with adequate human control \citep{ yang2020re}.  

\subsection{Privacy}
Privacy is a significant concern in all systems where the use of personal data has significant social and economic impact \citep{ji2014differential}. 
Concerns over privacy in AI systems are particularly prevalent, with the high volume of data used for sensitive decisions, such as advertisement, surveillance, health-care decisions, and money lending \citep{fjeld2020principled}. In the European Union, the General Data Protection Regulation (GDPR) highly regulates the use of personal and private data in AI, with large financial penalties levied on businesses who do not comply \citep{GDPR}. In particular, the clause that provides users with the "right to be forgotten", where users have the right to request that their personal data be deleted, is particularly important in AI privacy. Researchers have called attention to the issues with applying this clause to AI, pointing out that the complete deletion of private data in AI may be impossible as AI systems do not "forget" data in the same way as humans \cite{villaronga2018humans}.



\subsection{Professional Responsibility}
Professional Responsibility targets the individuals and entities involved in AI system design, development, and deployment \citep{fjeld2020principled}. As these individuals have a direct effect on the behavior and impacts of AI systems, it is vital for us to consider the intentions, abilities, integrity, and trustworthiness of such individuals. Research in this field focuses on developer responsibility (ethical, legal, and scientific) for the design and impact of their systems \citep{coeckelbergh2020artificial}. 

\subsection{Promotion of Human Values}
The promotion of human values (often also called the \textit{beneficence} principle \cite{floridi2019unified}) is of particular importance when considering the ethical implications of AI systems. This principle implies that AI should be designed and strongly influenced by human values, including moral, ethical, and societal norms \cite{dignum2017responsible}. This principle also includes ensuring that AI are leveraged to benefit society, aim toward positive change, and consider sustainability and environmental impact \cite{ec2019ethics}. This principle is very broad
and lacks a concrete definition within the field, making technical implementations challenging \citep{hagendorff2020ethics}. There are no known tools that deal directly with providing technical applications for human values-alignment during AI development \cite{morley2019overview}.

\subsection{Safety \& Security}
Safety and security are both vital to consider when developing trustworthy AI. In recent years, damages caused by autonomous vehicles, manipulation of public-facing AI systems, and software problems have harmed public perceptions of the safety and security of AI systems in society \citep{amodei2016concrete}. This principle covers AI safety, assessing the safety of AI systems, and security, assessing how secure an AI system is, and ensuring the robustness of an AI system from adversarial attacks. 

AI safety is both a technical and ethical concern, where potentially negative impacts on society could occur due to unintended accidents or failures \citep{varshney2017safety}.
Security flaws can contribute to these failures, where attacks by
malicious actors can misclassify inputs to worsen or manipulate performance or gain information about the model and data it was trained on \citep{ibitoye2019threat}. Often, the principles of privacy and safety and security are interconnected, where issues in one domain are likely to have an impact on the other. For example, \cite{liu2018survey} found that information leakage in the privacy domain affects model robustness and adversarial security. 

Several solutions have been proposed to improve the safety and security of AI systems. To achieve safety, one may consider the four principles of safety: Inherently safe design, Safety reserves, Safe fail and Procedural safeguards \cite{moller2008principles}. In regards to security, while no solution is complete, systems can be developed to detect and address adversarial issues and attacks \citep{nicolae2018adversarial}.  

\subsection{Transparency \& Explainability}
Transparency and explainability refer to the principle that the operations and outcomes of an AI model should be understandable to a human. This area of research is quite active, with many advocating for explainable decisions and transparency in sharing data and model information. These concepts are particularly vital to increasing trust in AI \citep{lipton2018mythos}. 

Research into explainability and transparency aims toward \textit{interpretablity}, developing AI in which a person can understand a model and its decisions, which in turn increases trust in the system \citep{doshi2017towards}.  At the base level, users should understand how a model is developed, its function, and how it reached its outcomes. This requires transparency. Ideally, developers should be transparent about an AI system's quality, intent, performance, and reasoning \citep{iyer2018transparency}.

However, as modeling becomes more complex, understanding becomes more difficult and opaque. Providing an interpretation for how an AI model works becomes a significant issue, as does providing a metric for measuring a model's explainability.  In addition, the type of explanation needed depends on the user, and therefore so does the metric needed to measure explainability, further complicating the issue \citep{hoffman2018metrics}.
The field of explainability and transparency is interdisciplinary and additional research is needed to formalize model interpretability, its evaluation, and how transparent creators should be about their models \citep{adadi2018peeking}. The field is very active, with solutions including utilizing transparent algorithms and models and providing post-hoc explanations about decisions. For a complete analysis of this field, we suggest the article by \citet{arrieta2020explainable}. 

\section{Related Work} \label{sec:related}
In recent years, there has been an increase in attempts to improve the transparency of the creation and deployment of AI. This includes transparency in model and data sharing, data lineage, and tracking the entire machine learning lifecycle. This section describes related work in tracking the development of AI systems, including a summary of the current state-of-the-art in AI provenance and transparency trends, data lineage, and machine learning lifecycle tracking. 


\subsection{Current Trends Toward AI Transparency}
Transparency in AI development and deployment requires clear communication of a variety of factors, such as the attainment and use of data, model development, deployment, and updating, and the functional details and purpose of systems. In particular, in recent years, importance has been placed on machine learning and data \textit{provenance}. Provenance in AI includes the collection and processing of information about the end-to-end development process of an AI system or use of data, aiming to improve replication, tracing, quality, and trust \cite{herschel2017survey}. 

\subsubsection{Data Transparency}
As the outcomes of AI systems depend directly on training data use (and misuse), data transparency, including transparency in data collection, utilization, and storage, is an area of significant concern in trustworthy AI. 

Data provenance (or data lineage) methods aim to improve replication, tracing, quality assessment in data use and data transformation processes \cite{herschel2017survey}. Several researchers have proposed data provenance and lineage solutions for the tracking of data and data transformations during the machine learning lifecycle \cite{zhang2017diagnosing, souza2019efficient, souza2019provenance}. Further, \citet{bertino2019data} proposes the use of blockchain technology to encourage data transparency and ensure that data collection and utilization coincide with ethical principles. 

While these solutions assist with internal data provenance, several researchers have also advocated for private, secure, and standardized methods for data sharing. \citet{gebru2018datasheets} proposed \textit{datasheets for datasets}, a standardization method for the documentation of datasets. These datasheets include information on "operating characteristics, test results, recommended uses, ... motivation, composition, collection process, [and] recommended uses", providing detailed questions for dataset creators to provide. Similarly, \citet{bender2018data} propose \textit{data statements} for dataset characterization in natural language processing, considering also the generalization of experiments and composition of datasets in respect to bias. Further, \citet{holland2018dataset} propose a standardized diagnostic method for an overview of the core components of a dataset with the \textit{dataset nutrition label}. 

Considering legality and regulations, \citet{yanisky2019equality} propose the \textit{AI Data Transparency Model}, encouraging data audits by both stakeholders and third-parties to assess data use (and misuse) and storage, to encourage replicability and compliance.

\subsubsection{Model Transparency}
Due to the rising complexity in modeling, model transparency and provenance methods have quickly gained traction. Research has focused both on end-to-end tracking of provenance information in the machine learning lifecycle, and in evaluation of models for performance and trust.

Several modeling provenance solutions have been proposed. \citet{schelter2017automatically} propose a system for the extraction and storage of meta-data and provenance information commonly observed in the machine learning lifecycle. \citet{hummer2019modelops} propose ModelOps, a cloud-based framework for end-to-end AI pipeline management, including support for addressing several trustworthy principles, such as reliability, traceability, quality control, and reproducibility. Further, several tools for complete asset tracking of AI pipelines have also been developed, focusing on tracking modeling inputs, results, and production processes \cite{zaharia2018accelerating, gharibi2021automated, idowu2021asset}. 


In regards to AI documentation, a recent trend is the use of \textit{FactSheets}. \citet{arnold2019factsheets} proposes FactSheets to communicate "purpose, performance, safety, security, and provenance information" from the creator to the user of an AI service. \citet{sokol2020explainability} extended this with a taxonomy for characterizing and assessing explainability in AI with \textit{Explainability FactSheets}. However, \citet{hind2020experiences} found that developers found these FactSheets challenging and time-consuming to complete, noting issues with developer recall about modeling details, data transformation documentation, privacy and ownership concerns, and lack of clarity. 


\subsection{Summary}
It is clear that academics and industry alike have established practices to encourage transparency and increase trust in AI development and deployment. The current focus is placed on data and model provenance, aiming to improve replicability, tracing, quality assessment, and trustworthiness in the AI lifecycle. 

While research has focused on tracking information about AI development, there are no concrete solutions for integrating transparent solutions with trustworthy principles. Current solutions focus primarily on one stage of the AI lifecycle, or only a handful of trustworthy principles, neglecting to give proper attention to the "whole picture" required in developing a trustworthy system. To increase trust in AI, we propose a framework that creators can leverage to increase the transparency and trustworthiness of their AI development processes.

\section{Know Your Model (KYM)} \label{sec:KYM}
In this section, we provide an overview of our proposed KYM framework. We propose 20 guidelines that provide clarity on the \textbf{efficacy}, \textbf{reliability}, \textbf{safety}, and \textbf{responsibility} of a given AI system's purpose, data treatment, modeling processes, and trustworthiness. 
These guidelines provide a framework for creators to leverage in their AI development processes to increase transparency and trustworthiness. This framework aims toward increasing user trust in AI systems and outcomes and easing the burden on creators by providing a clear set of guidelines
that considers provenance, trust, and technical, ethical, and legal responsibility.

The concept of KYM is influenced by the idea that all models have a unique identity and that model characteristics can be leveraged to know and trust models. To "know" a model implies collecting, recording, and storing detailed records of the processes undergone during the development of a model, subsequently establishing \textit{model identity}.
A developer should record enough information about a model to clearly establish its identity. In this case, model identity refers to the minimum information to distinguish one model from another, or establish a model's uniqueness. KYM strives for all models to have a unique model identity, allowing model characteristics to be leveraged to know and trust models. 

To encourage large-scale AI adoption, we must increase the trust that users have in models and modeling processes. The user may be a customer, employee, or the developer of the system, trust from each is equally important. While a solution to create a perfect trustworthy AI system is not currently available, by increasing transparency in how developers record how they address each principle, we believe that users will have the information needed to assess a model's trustworthiness. 

This need for transparency highlights the necessity of the KYM framework. KYM provides a framework for developers to address key areas about their AI systems, focusing on {efficacy}, reliability, safety, and responsibility. These four key concepts cover all eight principles of trustworthy AI to ensure complete coverage in KYM. 
Further, KYM includes guidelines for the utilization and preparation of data, modeling processes (model type, hyperparameter tuning, feature extraction, etc.), and methods for addressing aspects of trustworthiness in model development.



While the framework of KYM addresses all principles, it is the developers' responsibility to decide which guidelines are most important to address for their system. We advocate for transparency in the use of data, development of models, and how issues of AI trustworthiness are addressed, while recognizing that this transparency is different among the various types of AI systems.  As definitively proving that a model is trustworthy is quite difficult, we suggest that developers maintain thorough records on methods taken to address trust concerns. Developers should record techniques and tools they used to address issues in each area, and justifications for why a check was not completed or required. Due to the rapidly evolving nature of AI development and AI research, KYM suggests that developers remain vigilant in addressing issues of trust with regular checks and updates, particularly in respect to fairness, privacy, security, safety, user understanding, and reliability. 


The rest of this section outlines the four principles of KYM. In the following section, the guidelines for KYM are proposed. 

\subsection{Efficacy}
Efficacy in KYM ensures that models produce the desired result. With the increase in the use of AI in everyday settings, it is vital to ensure that the outcomes of models are appropriate for their intended purpose, that the model performs well, and that outcomes are fair and beneficial to society. As systems can have unintended outcomes, it should be verified that models perform in the way that the developer intended. 

The concept of Efficacy addresses the trustworthy AI principles of \textit{Transparency \& Explainability} and \textit{Fairness \& Non-Discrimination}. In KYM, the principle of efficacy calls for:
\begin{itemize}
    \item Transparency in the purpose, intentions, and outcomes of models, including intended purpose, use, target groups, and expected outputs.
    \item Efforts toward improving human understanding of processes, operations, and outcomes of the AI pipeline.
    \item Careful attention to fairness and non-discrimination in data and modeling to reduce bias and discrimination in outcomes.
\end{itemize}


\subsection{Reliability}
Reliability in KYM ensures that models are reliable in their outputs and developmental processes. The concept of reliability in KYM primarily addresses the trustworthy AI principle of \textit{Accountability}. Here, it is important to consider the processes that are used in development: Is the process appropriate for the intended purpose? Are the outcomes and processes verifiable, reproducible, and reliable? Would another method produce more reliable results? Are the appropriate regulatory and legal processes followed?

Of critical importance in this concept is replicability: developers should be able to reproduce the outcomes of their models and trace the model back to its origin. This includes ensuring proper provenance with records of data used, data transformations undergone, modeling processes (development environment, model type, hyperparameter tuning, etc.), and inference verification. Users should be able to verify the developmental products of models. KYM advocates that developers keep clear records of their model development so that a clear auditing process can be completed.

Additionally, attention should be given to data and modeling quality. The collection, preparation, and treatment of data are vital to consider when considering model identity. Data has a profound impact on the modeling process. The type and quality of data used for the development of an AI system have direct consequences on the quality of the models and inferences. For instance, data of poor quality or poorly leveraged data can lead to unreliable and incorrect models \citep{sanders2017garbage}. 
Therefore, careful attention to data processes is required. KYM requires reliability in the data use and data transformation in modeling systems.



The principle of reliability calls for: 
\begin{itemize}
    \item Transparency in developmental processes, including the use and transformation processes of data, and feature extraction, training and testing, and prediction outcomes. 
    \item Reliability in outcomes and developmental processes, including the appropriate use of methods, availability, and consistency.
    \item Replicability or verifiability of outcomes and processes.
    \item Attention to data quality to avoid bad, inadequate or inappropriate data collection, utilization, or transformation processes.
    \item Data and model provenance.
    \item Attention to ethical, legal, and regulatory environments and requirements.
\end{itemize}

This concept is particularly important in regulatory environments, where developers may be required to verify the exact processes undergone during model development and reproduce relevant results.



\subsection{Safety}
The large-scale adoption of AI requires that users are confident that AI systems are safe to use and do not pose undue harm to the user or society as a whole. The need for safety is considered with great importance in KYM. The concept of Safety in KYM addresses the trustworthy AI principles of \textit{Safety \& Security}  and \textit{Privacy}. Here, the concept of safety includes assessing the safety, security, and privacy of AI systems from unintended accidents, breaches, and threats to user privacy. 

This principle calls for:
\begin{itemize}
    \item Building AI systems with careful attention to safety, including safe design, contingencies in case of error or failure, and audits or standards to assess initial and continuous system safety.
    \item System and model stability, including attention to failures and their causes, maintenance to address and fix failures upon occurrence, and reducing failure rates \cite{saria2019tutorial}.
    \item Robustness to threats to security, including robustness to attacks from adversaries or malicious actors and continual attention to state-of-the-art security techniques. 
    \item Careful attention to user privacy, including (personal) data collection, utilization, and storage. This also includes any legal or regulatory requirements for securing user information.
\end{itemize}



\subsection{Responsibility}
Responsibility in KYM bridges the gap between technical implementation and legal and ethical implications, addressing the trustworthy AI principles of \textit{Professional Responsibility}, \textit{Human Control of Technology}, and \textit{Promotion of Human Values}. In addition to technical information about AI systems, creators must pay close attention to societal, social, and developer roles in the overarching impact of their systems. 

The principle of responsibility in KYM is perhaps the most abstract. 
With the large variation in the applications of AI systems, responsibility will have a different meaning to each creator. Rather than providing concrete guidelines in this area, KYM encourages creators to be transparent about the impacts and purposes of their systems, who was involved in their creation, and the level at which human control is required and provided.


The principle of responsibility calls for: 

\begin{itemize}
    \item Transparency about developer or creator identity, including transparency about stakeholders and entities involved in the design and deployment of AI systems. 
    \item Careful attention to the level at which human control is required and provided, including clarity on the implementation of human control in a system, opportunities for human intervention and review, and safeguards in the absence of human control.
    \item Consideration of the societal impact, purpose, and value of AI systems, and methods to maximize their benefit to society.
\end{itemize}

\newcolumntype{b}{>{\hsize=1.3\hsize}X}
\newcolumntype{s}{>{\hsize=.7\hsize}X}

\section{Key Guidelines of KYM} \label{sec:requirements}
\newcommand\RotText[1]{\rotatebox{90}{\parbox{3.5cm}{\centering#1}}}

\begin{table*}[]
    \centering
        \caption{The Know Your Model (KYM) Guidelines for Efficacy (E\#) and Reliability (RL\#). An example application of each guideline is provided. These examples are simplified. Real-world applications may require longer and more technical justifications.}
    \begin{tabularx}{\linewidth}{lsb} 
         \multicolumn{2}{l}{Know Your Model Guidelines}  & Example Application of Guideline \\
         \midrule
                    E1 & \textbf{Creators SHOULD describe the intended purpose, use, target user, and output of the system.} & [Navigation] "The system is a navigation system that users can use to map the most efficient path from one location to another. The system outputs the shortest path as defined by the estimated travel time from one input to another, utilizing available geographical information at the time of request." 
         \\
         E2 & \textbf{Creators MUST record (statistical) metrics about training and test datasets.} & [Logistics AI] "The system was trained on a combination of our weekly, quarterly, and annual volume information. This data shows an average purchase of 10,000 units.(sd = 1,000), with higher throughput events with an average of 15,000 units (sd=2,500) occurring around holidays. It was confirmed that the training and test datasets exhibit identical distributions."  \\
         E3 & \textbf{Creators SHOULD describe the expected performance on unseen data} &  [Medical AI] "Data from low-quality or outdated equipment will result in poor performance. Shadowing or blurring in images may negatively affect model performance."  \\ 
         E4 &  \textbf{Creators SHOULD record \textit{methods} taken to reduce bias, discrimination, and fairness issues in data and modeling outcomes, and SHOULD record specific \textit{metrics} on bias, discrimination, and fairness.} & [Criminal Sentencing AI] "In order to ensure racial fairness in sentences, all potentially identifying racial information has been removed from the dataset. Additionally, the system was evaluated by experts in racial justice and equality in order to mitigate potential problems with bias. Bias remediation was performed using [state-of-the-art tool]. A bias was identified and mitigated with a re-weighing method." \\ 
         E5 & \textbf{Creators SHOULD aim for increased understandability.} 
         & [Medical AI] "This model is designed to be used by trained doctors. The system provides diagnostic information and justifications that explain the qualities used for each decision. Furthermore, the system provides documentation to provide additional information."\\

         RL1 & \textbf{Creators MUST record the processes followed in the development of the AI system.} 
         &  [E-Commerce AI] "This model leverages neural network technology, building on research previously published in the domain. Model training and testing were tracked locally and will be stored for three years following the end-of-life of the product. Data is collected and stored in accordance with international regulation." \\
         RL2 & \textbf{Creators MUST ensure adequate provenance for data.} 
         & [Social Media AI] "Textual data was parsed from three social media websites between the dates of January and May 2020, and stored on a private server. Data were not checked for quality. Datasets are documented internally. The system maintains an index of all data as well as a log of all changes to the data set. Unigram transformation and punctuation removal were utilized."\\
         RL3 & \textbf{Creators MUST ensure adequate provenance for end-to-end model development.}  & [Advertising AI] "Complete records of metadata from model training, testing, and prediction were taken utilizing an end-to-end asset tracking tool."\\
         
         RL4 & \textbf{Creators MUST record evaluation and performance metrics.} & [Classification AI] "Models were trained using a 70/30 test/train split, 10-fold cross-validation, and evaluated using prediction accuracy and AUC. The chosen model has an 80.2\% accuracy rate, with a sensitivity/specificity rate of 74.5\%/61.8\% respectively." \\
         RL5 & \textbf{Creators SHOULD track model update performance and information ingestion.} 
         & [Social Media AI] "We capture user data upon each deployment and retrain the model with the captured data. Model performance is analyzed with each update and must remain within $\pm 15\%$." \\
         RL6 & \textbf{Creators SHOULD record metrics on outcome replicability.} & [Robotics AI] "In order to reproduce the system results, a docker file has been provided. By leveraging this dataset and docker file, the system will produce the same results. This docker file was created using the following dataset and model settings."  \\
    \end{tabularx}
    \label{tab:tab:ER}
\end{table*}

\begin{table*}[]
    \centering
        \caption{The Know Your Model (KYM) Guidelines for Safety (S\#) and Responsibility (RS\#). An example application of each guideline is provided. These examples are simplified. Real-world applications may require longer and more technical justifications.}
    \begin{tabularx}{\linewidth}{lsb} 
         \multicolumn{2}{l}{Know Your Model Guidelines}  & Example Application of Guideline \\
         \midrule

         S1 & \textbf{Creators MUST assess safety to users and society.} 
         & [Robotics AI] "In the event of detected compromise, the system can be placed into a fail-safe state by the activation of a hardware cutoff or a software shutdown. In order to comply with safety standards, this system has several human-tracking safety features that override the AI in situations where humans can potentially be harmed." \\
         S2 & \textbf{Creators MUST assess potential security, safety, and privacy failure points.} & [Finance AI] "The system was designed with the following threat model in mind. The system is an online banking platform with the potential for both denial-of-service, and database attacks. Additionally, the model is trained on user-data which has been anonymized, however, attacks do exist that could de-anonymize users. Finally, the model itself is vulnerable to data poisoning or similar attacks.   " \\
         S3 & \textbf{Creators SHOULD record metrics for security robustness.} & [E-Commerce AI] "Our system is regularly tested to comply with PCI DSS standards. We have also received ISO/IEC 27001:2013 certification for our handling of critical data."  \\
         S4 & \textbf{Creators MUST ensure user privacy, and appropriate treatment and use of private data.} & [E-Commerce AI] "Only data that is relevant to the product is collected, with consent of the individual. Private data is stored on an encrypted server." \\
         S5 & \textbf{Creators SHOULD ensure secure data utilization and storage.}  & [Personal Services AI] "Data is stored on an encrypted disk, where access is granted by keys. All data changes are signed by key, for easy traceability." \\
         
         RP1 & \textbf{Creators SHOULD disclose or record all entities involved in system development.} 
         & [Human Resource AI] "Our team is composed of machine learning engineers, statisticians, and social scientists, all graduates of accredited universities. We consulted with an AI domain expert during development." \\
         RP2 & \textbf{Creators SHOULD detail the implementation of human-AI interactions.} & [Medical AI] "The system uses patient characteristics and health information to formulate diagnoses. The decisions must be confirmed by a human before a diagnosis can be made."\\
         RP3 & \textbf{Creators SHOULD describe the impact, value, and benefit of the system.} & [Chatbot] "The system allows for rapid interactions with customers. This increases availability, provides immediate assistance to customers, and reduces the need for customer service staff. The system is only used for our business and does not have any larger foreseen societal impacts." \\
         RP4 & \textbf{Creators MUST comply with legal and regulatory requirements.} & [Finance AI] "Our system complies with GDPR regulations on the use of private data, and internal regulations on the use of private data and clarity in decisions." \\
    \end{tabularx}
    \label{tab:tab:SR}
\end{table*}

In this section, we define the key guidelines of KYM. These guidelines summarize the information developers are encouraged to record to establish model identity. The key words "MUST", "MUST NOT", "REQUIRED", "SHALL", "SHALL
      NOT", "SHOULD", "SHOULD NOT", "RECOMMENDED",  "MAY", and
      "OPTIONAL" in this document are to be interpreted as described in
      RFC 2119 \cite{bradner1997rfc2119}.

Example applications of each of these guidelines are provided in Tables \ref{tab:tab:ER} and \ref{tab:tab:SR} .


\subsection{Efficacy}
\subsubsection{E1: Creators SHOULD describe the intended purpose, use, target user, and outputs of the system.} 
Creators SHOULD record information on the intentions of their AI systems. This may include brief descriptions of the intended purpose or goals of the system, expected use, sample use-cases, target user of the system and its outcomes, release dates, and time-frame-of-validity of the system.   
In cases where intentions and outcomes are misaligned, the extent of the misalignment and any positive and negative impacts SHOULD be known and recorded. Further, Creators SHOULD (briefly) be transparent about the expected outputs of the system, as the expected and observed outputs may differ.


\subsubsection{E2: Creators MUST record (statistical) metrics about training and test datasets.} To ensure that the training and test dataset distributions match, metrics about the datasets MUST be recorded. If applicable, this SHOULD include metrics on demographic information. 


\subsubsection{E3: Creators SHOULD describe the expected performance on unseen data.} 
Once deployed, AI systems may experience data that is vastly different than the data used to train/test the system. Creators SHOULD describe the expected performance on unseen data, such as data from different distributions. 

\subsubsection{E4: Creators SHOULD record \textit{methods} taken to reduce bias, discrimination, and fairness issues in data and modeling outcomes, and SHOULD record specific \textit{metrics} on bias, discrimination, and fairness..} 
Creators SHOULD record any methods taken to address bias, discrimination, and fairness issues in data or modeling outcomes. This may include data treatment techniques and remediation, model checks and remediation, and outcome verification.

Even in cases where careful attention is paid to reduce bias in input data, algorithms may still exhibit biased behaviors. Developers are encouraged to pursue methods to measure fairness in their outcomes, using state-of-the-art methods and tools.


\subsubsection{E5: Creators SHOULD aim for increased understandability.} Developers SHOULD attempt to increase understanding of all stages of AI development to different users and groups. Detail efforts taken to improve explainability, transparency, human-AI interactions (review, validation, etc.) of the developmental processes and outcomes of AI systems. This may include providing explanations on model decisions, clarity in model processes and techniques utilized, and interpreting model development and functionality in language appropriate to the target user. 

\subsection{Reliability}

\subsubsection{RL1: Creators MUST record the processes followed in the development of the AI system.} Document and justify the implemented algorithms and techniques, collection, utilization, and storage of data, verification and testing methods, and output generation of the system. Documentation MUST be thorough and include all information needed to identify and justify utilized methods, identify storage locations, and replicate outcomes. The extent of the required information will vary greatly depending on system type. 

\subsubsection{RL2: Creators MUST ensure adequate provenance for data.} Creators MUST maintain clear records of data collection, utilization, and transformation processes. Records MUST be adequate, clear, and complete enough to determine the origin of the data, assess data quality, and understand any transformations that occurred. Records may include, but are not limited to, data collection process and techniques, the identity of data owner or licensing entity, dataset creation time, type and amount of data utilized, dataset utilization in development, and data updating practices. 


\subsubsection{RL3: Creators MUST ensure adequate provenance for end-to-end model development.} Developers MUST maintain clear records of developmental processes undergone in AI design, development, and deployment. These records MUST be complete enough to be able to replicate model results and outcomes. Records may include data (and/or meta-data) on feature extraction, training and testing, and prediction outcomes, date and time of modeling stages, development environment (development language, packages used, etc.), model version, time of the last update, changes in performance between updates, algorithms and techniques used, training conditions (i.e. hyperparameters), use of the dataset in each stage, testing performance \& results, etc. 

\subsubsection{RL4: Creators MUST record evaluation and performance metrics} Developers MUST record detailed records of the evaluation and performance processes used. Creators MUST maintain a record of the metrics and techniques that were used to measure the performance of their systems, such as accuracy, precision/recall, error rates, F-1 scores, AUC, etc.  It is suggested that significant technical data is recorded. Metrics for both intermediary and final models are encouraged.


\subsubsection{RL5: Creators SHOULD track model update performance and information ingestion.} Developers SHOULD clearly track model updates and how new data is used and affects performance. If new data is ingested after deployment, developers SHOULD record the origin of the new data, how it is integrated into the system, and if there are any bounds for performance changes.

\subsubsection{RL6: Creators SHOULD record metrics on outcome replicability.}
Developers SHOULD measure the replicability of outcomes of their AI systems, utilizing state-of-the-art metrics.

\subsection{Safety}
\subsubsection{S1: Creators MUST assess safety to users and society.} The development of systems MUST consider safety at the forefront. Developers MUST pay careful attention to safe design, failure contingencies, and safety standards. Consideration MUST be given to how the AI system impacts its surroundings, individuals, and society as a whole, and whether its use or deployment poses any safety risks. In the case that there are safety concerns, creators MUST be transparent in any safety concerns or issues the AI system may have.

\subsubsection{S2: Creators MUST assess potential security, safety, and privacy failure points.} Assessments of potential security, safety, and privacy failure points present in models (and solutions if available) MUST be undertaken. 

\subsubsection{S3: Creators SHOULD record metrics for security robustness.} Creators SHOULD record metrics taken for improving the robustness of their systems from adversarial attacks and malicious actors (i.e. checks undergone for adversarial concerns). Due to the rapidly evolving nature of AI security, developers SHOULD continuously engage in improving security robustness utilizing state-of-the-art techniques.  

\subsubsection{S4: Creators MUST ensure user privacy, and appropriate treatment and use of private data.} Developers MUST be acutely aware of the treatment of user data and the role of user data in their systems development and outcomes. For private data, creators SHOULD consider regulatory requirements for storage, deletion, and use of data, including requirements for consent.

\subsubsection{S5: Creators SHOULD ensure secure data utilization and storage.} Creators SHOULD ensure that all data is used and stored securely.

\subsection{Responsibility}
\subsubsection{RP1: Creators SHOULD disclose or record all entities involved in system development.} Record the identities (or affiliations), qualifications, and diversity of all entities involved (including stakeholders, businesses, domain experts, individuals, teams, etc.) in the design, development, and deployment of the AI system. This may include the experience and credentials of developers, team diversity, and the investments and interests of developers (and other stakeholders) in model development.

\subsubsection{RP2: Creators SHOULD detail the implementation of human-AI interactions.} Creators SHOULD understand the implementation of human-AI interactions in the system. This may include areas where human review is allowed and/or required, opportunities for human intervention, and human role in AI decisions. 

\subsubsection{RP3: Creators SHOULD describe the impact, value, and benefit of the system.} Creators SHOULD justify the impacts, values, and benefits that the AI system has to society. This may also include any potential detriments to society (and justifications for why the AI system maintains value).

\subsubsection{RP4: Creators MUST comply with legal and regulatory requirements.}
With the rising legal and regulatory requirements for AI development, careful attention MUST be given to national, international, and vocational requirements for AI design, development, and deployment.

\section{Discussion and Future Work} \label{sec:discussion}


With the proliferation of AI into greater society, members of academia and industry alike should strive for the development of robust, trustworthy systems. The complexity and wide array of applications in AI systems complicate the process of creating trust, placing the burden on each creator to establish a method for building and maintaining trust. Toward developing trust in AI, we propose the Know Your Model (KYM) framework, a set of guidelines that can be leveraged to establish model identity and increase transparency and trust in their AI development processes.

The KYM guidelines aim to provide a comprehensive framework for creators to leverage to address both provenance and principles of trust in the design, development, and deployment of their AI systems. Although previous efforts have been made to increase transparency and trust in AI, the focus has been placed primarily on provenance rather than trust. In those methods that do address trust, attention is only given to one or two principles, neglecting the importance of others. Our framework aims to merge provenance methods with a focus on trust, providing a complete framework for creators to assess their current and future AI processes. Further, our framework considers the importance of technical, ethical, and legal responsibility, providing guidelines that bridge the gap between research and industry.

As definitively proving that a system or model is trustworthy is quite difficult, we suggest that developers maintain thorough records on methods taken to address trust concerns. By increasing transparency, developers ensure clarity on how key issues are addressed, and users have the information needed to assess trust where necessary. Areas of trust to address include the eight principles of trustworthy AI: accountability, fairness and non-discrimination, human control, privacy, professional responsibility, promotion of human values, safety and security, and transparency and explainability. Developers should record techniques and tools they used to address issues in each area, and justifications for why a check was not completed or required. These should be frequently checked and updated for every new model or model update. Given the current state of trust in AI, we believe that increasing transparency in this way is the next step in increasing overall trust.  

It should be noted that while KYM provides a set of guidelines for creators to leverage, it does not provide a method for \textit{sharing} this information among users, or outside of an organization. This distinction should be made by the creator depending on the unique factors of their systems. In general, we advocate for thorough records and complete transparency within the development team and internal users (where applicable), but understand that information sharing and transparency will be considerably different with external users, such as customers and the general public. 


We believe that further attention is warranted on developing a formalized system for KYM. As the state of AI research is rapidly evolving, we do not suggest specific methodologies to address the KYM guidelines technically (such as specific tools or software to analyze fairness or bias). It would be of benefit to develop a formalized system which includes up-to-date methods to analyze the guidelines technically. Further, it would be of extreme value if the guidelines of KYM could be streamlined into an automated system for record-keeping for creators to leverage. Future work in this area is needed. This may include clear avenues for the sharing of information with external users, such as customers or the general public.


It is our hope that KYM, model identity, and transparency in AI become more prevalent over time. With KYM, we strive to provide a framework to guide creators in increasing transparency and trust in AI. With these guidelines, we hope to encourage a new era of transparency and trust in AI. 

\bibliographystyle{ACM-Reference-Format}
\bibliography{uai2021-template}


\begin{thebibliography}{51}


\ifx \showCODEN    \undefined \def \showCODEN     #1{\unskip}     \fi
\ifx \showDOI      \undefined \def \showDOI       #1{#1}\fi
\ifx \showISBNx    \undefined \def \showISBNx     #1{\unskip}     \fi
\ifx \showISBNxiii \undefined \def \showISBNxiii  #1{\unskip}     \fi
\ifx \showISSN     \undefined \def \showISSN      #1{\unskip}     \fi
\ifx \showLCCN     \undefined \def \showLCCN      #1{\unskip}     \fi
\ifx \shownote     \undefined \def \shownote      #1{#1}          \fi
\ifx \showarticletitle \undefined \def \showarticletitle #1{#1}   \fi
\ifx \showURL      \undefined \def \showURL       {\relax}        \fi
\providecommand\bibfield[2]{#2}
\providecommand\bibinfo[2]{#2}
\providecommand\natexlab[1]{#1}
\providecommand\showeprint[2][]{arXiv:#2}

\bibitem[\protect\citeauthoryear{Adadi and Berrada}{Adadi and Berrada}{2018}]%
        {adadi2018peeking}
\bibfield{author}{\bibinfo{person}{Amina Adadi} {and} \bibinfo{person}{Mohammed
  Berrada}.} \bibinfo{year}{2018}\natexlab{}.
\newblock \showarticletitle{Peeking inside the black-box: a survey on
  explainable artificial intelligence (XAI)}.
\newblock \bibinfo{journal}{\emph{IEEE access}}  \bibinfo{volume}{6}
  (\bibinfo{year}{2018}), \bibinfo{pages}{52138--52160}.
\newblock


\bibitem[\protect\citeauthoryear{Amodei, Olah, Steinhardt, Christiano,
  Schulman, and Man{\'e}}{Amodei et~al\mbox{.}}{2016}]%
        {amodei2016concrete}
\bibfield{author}{\bibinfo{person}{Dario Amodei}, \bibinfo{person}{Chris Olah},
  \bibinfo{person}{Jacob Steinhardt}, \bibinfo{person}{Paul Christiano},
  \bibinfo{person}{John Schulman}, {and} \bibinfo{person}{Dan Man{\'e}}.}
  \bibinfo{year}{2016}\natexlab{}.
\newblock \showarticletitle{Concrete problems in AI safety}.
\newblock \bibinfo{journal}{\emph{arXiv preprint arXiv:1606.06565}}
  (\bibinfo{year}{2016}).
\newblock


\bibitem[\protect\citeauthoryear{Arnold, Bellamy, Hind, Houde, Mehta,
  Mojsilovi{\'c}, Nair, Ramamurthy, Olteanu, Piorkowski, et~al\mbox{.}}{Arnold
  et~al\mbox{.}}{[n.d.]}]%
        {arnold2019factsheets}
\bibfield{author}{\bibinfo{person}{Matthew Arnold}, \bibinfo{person}{Rachel~KE
  Bellamy}, \bibinfo{person}{Michael Hind}, \bibinfo{person}{Stephanie Houde},
  \bibinfo{person}{Sameep Mehta}, \bibinfo{person}{Aleksandra Mojsilovi{\'c}},
  \bibinfo{person}{Ravi Nair}, \bibinfo{person}{K~Natesan Ramamurthy},
  \bibinfo{person}{Alexandra Olteanu}, \bibinfo{person}{David Piorkowski},
  {et~al\mbox{.}}} \bibinfo{year}{[n.d.]}\natexlab{}.
\newblock \showarticletitle{FactSheets: Increasing trust in AI services through
  supplier's declarations of conformity}.
\newblock \bibinfo{journal}{\emph{IBM Journal of Research and Development}}
  (\bibinfo{year}{[n.\,d.]}).
\newblock


\bibitem[\protect\citeauthoryear{Arrieta, D{\'\i}az-Rodr{\'\i}guez, Del~Ser,
  Bennetot, Tabik, Barbado, Garc{\'\i}a, Gil-L{\'o}pez, Molina, Benjamins,
  et~al\mbox{.}}{Arrieta et~al\mbox{.}}{2020}]%
        {arrieta2020explainable}
\bibfield{author}{\bibinfo{person}{Alejandro~Barredo Arrieta},
  \bibinfo{person}{Natalia D{\'\i}az-Rodr{\'\i}guez}, \bibinfo{person}{Javier
  Del~Ser}, \bibinfo{person}{Adrien Bennetot}, \bibinfo{person}{Siham Tabik},
  \bibinfo{person}{Alberto Barbado}, \bibinfo{person}{Salvador Garc{\'\i}a},
  \bibinfo{person}{Sergio Gil-L{\'o}pez}, \bibinfo{person}{Daniel Molina},
  \bibinfo{person}{Richard Benjamins}, {et~al\mbox{.}}}
  \bibinfo{year}{2020}\natexlab{}.
\newblock \showarticletitle{Explainable Artificial Intelligence (XAI):
  Concepts, taxonomies, opportunities and challenges toward responsible AI}.
\newblock \bibinfo{journal}{\emph{Information Fusion}}  \bibinfo{volume}{58}
  (\bibinfo{year}{2020}), \bibinfo{pages}{82--115}.
\newblock


\bibitem[\protect\citeauthoryear{Beede, Baylor, Hersch, Iurchenko, Wilcox,
  Ruamviboonsuk, and Vardoulakis}{Beede et~al\mbox{.}}{2020}]%
        {beede2020human}
\bibfield{author}{\bibinfo{person}{Emma Beede}, \bibinfo{person}{Elizabeth
  Baylor}, \bibinfo{person}{Fred Hersch}, \bibinfo{person}{Anna Iurchenko},
  \bibinfo{person}{Lauren Wilcox}, \bibinfo{person}{Paisan Ruamviboonsuk},
  {and} \bibinfo{person}{Laura~M Vardoulakis}.}
  \bibinfo{year}{2020}\natexlab{}.
\newblock \showarticletitle{A human-centered evaluation of a deep learning
  system deployed in clinics for the detection of diabetic retinopathy}. In
  \bibinfo{booktitle}{\emph{Proceedings of the 2020 CHI conference on human
  factors in computing systems}}. \bibinfo{pages}{1--12}.
\newblock


\bibitem[\protect\citeauthoryear{Bellamy, Dey, Hind, Hoffman, Houde, Kannan,
  Lohia, Martino, Mehta, Mojsilovic, et~al\mbox{.}}{Bellamy
  et~al\mbox{.}}{2018}]%
        {bellamy2018ai}
\bibfield{author}{\bibinfo{person}{Rachel~KE Bellamy}, \bibinfo{person}{Kuntal
  Dey}, \bibinfo{person}{Michael Hind}, \bibinfo{person}{Samuel~C Hoffman},
  \bibinfo{person}{Stephanie Houde}, \bibinfo{person}{Kalapriya Kannan},
  \bibinfo{person}{Pranay Lohia}, \bibinfo{person}{Jacquelyn Martino},
  \bibinfo{person}{Sameep Mehta}, \bibinfo{person}{Aleksandra Mojsilovic},
  {et~al\mbox{.}}} \bibinfo{year}{2018}\natexlab{}.
\newblock \showarticletitle{AI Fairness 360: An extensible toolkit for
  detecting, understanding, and mitigating unwanted algorithmic bias}.
\newblock \bibinfo{journal}{\emph{arXiv preprint arXiv:1810.01943}}
  (\bibinfo{year}{2018}).
\newblock


\bibitem[\protect\citeauthoryear{Bender and Friedman}{Bender and
  Friedman}{2018}]%
        {bender2018data}
\bibfield{author}{\bibinfo{person}{Emily~M Bender} {and} \bibinfo{person}{Batya
  Friedman}.} \bibinfo{year}{2018}\natexlab{}.
\newblock \showarticletitle{Data statements for natural language processing:
  Toward mitigating system bias and enabling better science}.
\newblock \bibinfo{journal}{\emph{Transactions of the Association for
  Computational Linguistics}}  \bibinfo{volume}{6} (\bibinfo{year}{2018}),
  \bibinfo{pages}{587--604}.
\newblock


\bibitem[\protect\citeauthoryear{Bertino, Kundu, and Sura}{Bertino
  et~al\mbox{.}}{2019}]%
        {bertino2019data}
\bibfield{author}{\bibinfo{person}{Elisa Bertino}, \bibinfo{person}{Ahish
  Kundu}, {and} \bibinfo{person}{Zehra Sura}.} \bibinfo{year}{2019}\natexlab{}.
\newblock \showarticletitle{Data transparency with blockchain and AI ethics}.
\newblock \bibinfo{journal}{\emph{Journal of Data and Information Quality
  (JDIQ)}} \bibinfo{volume}{11}, \bibinfo{number}{4} (\bibinfo{year}{2019}),
  \bibinfo{pages}{1--8}.
\newblock


\bibitem[\protect\citeauthoryear{Bradner}{Bradner}{1997}]%
        {bradner1997rfc2119}
\bibfield{author}{\bibinfo{person}{Scott Bradner}.}
  \bibinfo{year}{1997}\natexlab{}.
\newblock \bibinfo{title}{RFC2119: Key words for use in RFCs to Indicate
  Requirement Levels}.
\newblock
\newblock


\bibitem[\protect\citeauthoryear{Coeckelbergh}{Coeckelbergh}{2020}]%
        {coeckelbergh2020artificial}
\bibfield{author}{\bibinfo{person}{Mark Coeckelbergh}.}
  \bibinfo{year}{2020}\natexlab{}.
\newblock \showarticletitle{Artificial intelligence, responsibility
  attribution, and a relational justification of explainability}.
\newblock \bibinfo{journal}{\emph{Science and engineering ethics}}
  \bibinfo{volume}{26}, \bibinfo{number}{4} (\bibinfo{year}{2020}),
  \bibinfo{pages}{2051--2068}.
\newblock


\bibitem[\protect\citeauthoryear{Dignum}{Dignum}{2017}]%
        {dignum2017responsible}
\bibfield{author}{\bibinfo{person}{Virginia Dignum}.}
  \bibinfo{year}{2017}\natexlab{}.
\newblock \showarticletitle{Responsible artificial intelligence: designing AI
  for human values}.
\newblock  (\bibinfo{year}{2017}).
\newblock


\bibitem[\protect\citeauthoryear{Doshi-Velez and Kim}{Doshi-Velez and
  Kim}{2017}]%
        {doshi2017towards}
\bibfield{author}{\bibinfo{person}{Finale Doshi-Velez} {and}
  \bibinfo{person}{Been Kim}.} \bibinfo{year}{2017}\natexlab{}.
\newblock \showarticletitle{Towards a rigorous science of interpretable machine
  learning}.
\newblock \bibinfo{journal}{\emph{arXiv preprint arXiv:1702.08608}}
  (\bibinfo{year}{2017}).
\newblock


\bibitem[\protect\citeauthoryear{Doshi-Velez, Kortz, Budish, Bavitz, Gershman,
  O'Brien, Scott, Schieber, Waldo, Weinberger, et~al\mbox{.}}{Doshi-Velez
  et~al\mbox{.}}{2017}]%
        {doshi2017accountability}
\bibfield{author}{\bibinfo{person}{Finale Doshi-Velez}, \bibinfo{person}{Mason
  Kortz}, \bibinfo{person}{Ryan Budish}, \bibinfo{person}{Chris Bavitz},
  \bibinfo{person}{Sam Gershman}, \bibinfo{person}{David O'Brien},
  \bibinfo{person}{Kate Scott}, \bibinfo{person}{Stuart Schieber},
  \bibinfo{person}{James Waldo}, \bibinfo{person}{David Weinberger},
  {et~al\mbox{.}}} \bibinfo{year}{2017}\natexlab{}.
\newblock \showarticletitle{Accountability of AI under the law: The role of
  explanation}.
\newblock \bibinfo{journal}{\emph{arXiv preprint arXiv:1711.01134}}
  (\bibinfo{year}{2017}).
\newblock


\bibitem[\protect\citeauthoryear{Fjeld, Achten, Hilligoss, Nagy, and
  Srikumar}{Fjeld et~al\mbox{.}}{2020}]%
        {fjeld2020principled}
\bibfield{author}{\bibinfo{person}{Jessica Fjeld}, \bibinfo{person}{Nele
  Achten}, \bibinfo{person}{Hannah Hilligoss}, \bibinfo{person}{Adam Nagy},
  {and} \bibinfo{person}{Madhulika Srikumar}.} \bibinfo{year}{2020}\natexlab{}.
\newblock \showarticletitle{Principled artificial intelligence: Mapping
  consensus in ethical and rights-based approaches to principles for AI}.
\newblock \bibinfo{journal}{\emph{Berkman Klein Center Research Publication}}
  \bibinfo{number}{2020-1} (\bibinfo{year}{2020}).
\newblock


\bibitem[\protect\citeauthoryear{Floridi and Cowls}{Floridi and Cowls}{2019}]%
        {floridi2019unified}
\bibfield{author}{\bibinfo{person}{Luciano Floridi} {and} \bibinfo{person}{Josh
  Cowls}.} \bibinfo{year}{2019}\natexlab{}.
\newblock \showarticletitle{A unified framework of five principles for AI in
  society}.
\newblock \bibinfo{journal}{\emph{Issue 1.1, Summer 2019}} \bibinfo{volume}{1},
  \bibinfo{number}{1} (\bibinfo{year}{2019}).
\newblock


\bibitem[\protect\citeauthoryear{Gebru, Morgenstern, Vecchione, Vaughan,
  Wallach, Daum{\'e}~III, and Crawford}{Gebru et~al\mbox{.}}{2018}]%
        {gebru2018datasheets}
\bibfield{author}{\bibinfo{person}{Timnit Gebru}, \bibinfo{person}{Jamie
  Morgenstern}, \bibinfo{person}{Briana Vecchione},
  \bibinfo{person}{Jennifer~Wortman Vaughan}, \bibinfo{person}{Hanna Wallach},
  \bibinfo{person}{Hal Daum{\'e}~III}, {and} \bibinfo{person}{Kate Crawford}.}
  \bibinfo{year}{2018}\natexlab{}.
\newblock \showarticletitle{Datasheets for datasets}.
\newblock \bibinfo{journal}{\emph{arXiv preprint arXiv:1803.09010}}
  (\bibinfo{year}{2018}).
\newblock


\bibitem[\protect\citeauthoryear{Gharibi, Walunj, Nekadi, Marri, and
  Lee}{Gharibi et~al\mbox{.}}{2021}]%
        {gharibi2021automated}
\bibfield{author}{\bibinfo{person}{Gharib Gharibi}, \bibinfo{person}{Vijay
  Walunj}, \bibinfo{person}{Raju Nekadi}, \bibinfo{person}{Raj Marri}, {and}
  \bibinfo{person}{Yugyung Lee}.} \bibinfo{year}{2021}\natexlab{}.
\newblock \showarticletitle{Automated end-to-end management of the modeling
  lifecycle in deep learning}.
\newblock \bibinfo{journal}{\emph{Empirical Software Engineering}}
  \bibinfo{volume}{26}, \bibinfo{number}{2} (\bibinfo{year}{2021}),
  \bibinfo{pages}{1--33}.
\newblock


\bibitem[\protect\citeauthoryear{Hagendorff}{Hagendorff}{2020}]%
        {hagendorff2020ethics}
\bibfield{author}{\bibinfo{person}{Thilo Hagendorff}.}
  \bibinfo{year}{2020}\natexlab{}.
\newblock \showarticletitle{The ethics of AI ethics: An evaluation of
  guidelines}.
\newblock \bibinfo{journal}{\emph{Minds and Machines}} \bibinfo{volume}{30},
  \bibinfo{number}{1} (\bibinfo{year}{2020}), \bibinfo{pages}{99--120}.
\newblock


\bibitem[\protect\citeauthoryear{Herschel, Diestelk{\"a}mper, and
  Lahmar}{Herschel et~al\mbox{.}}{2017}]%
        {herschel2017survey}
\bibfield{author}{\bibinfo{person}{Melanie Herschel}, \bibinfo{person}{Ralf
  Diestelk{\"a}mper}, {and} \bibinfo{person}{Houssem~Ben Lahmar}.}
  \bibinfo{year}{2017}\natexlab{}.
\newblock \showarticletitle{A survey on provenance: What for? What form? What
  from?}
\newblock \bibinfo{journal}{\emph{The VLDB Journal}} \bibinfo{volume}{26},
  \bibinfo{number}{6} (\bibinfo{year}{2017}), \bibinfo{pages}{881--906}.
\newblock


\bibitem[\protect\citeauthoryear{{High-Level Expert Group on AI}}{{High-Level
  Expert Group on AI}}{2019}]%
        {ec2019ethics}
\bibfield{author}{\bibinfo{person}{{High-Level Expert Group on AI}}.}
  \bibinfo{year}{2019}\natexlab{}.
\newblock \bibinfo{booktitle}{\emph{Ethics guidelines for trustworthy AI}}.
\newblock \bibinfo{type}{Report}. \bibinfo{institution}{European Commission},
  \bibinfo{address}{Brussels}.
\newblock
\urldef\tempurl%
\url{https://ec.europa.eu/digital-single-market/en/news/ethics-guidelines-trustworthy-ai}
\showURL{%
\tempurl}


\bibitem[\protect\citeauthoryear{Hind, Houde, Martino, Mojsilovic, Piorkowski,
  Richards, and Varshney}{Hind et~al\mbox{.}}{2020}]%
        {hind2020experiences}
\bibfield{author}{\bibinfo{person}{Michael Hind}, \bibinfo{person}{Stephanie
  Houde}, \bibinfo{person}{Jacquelyn Martino}, \bibinfo{person}{Aleksandra
  Mojsilovic}, \bibinfo{person}{David Piorkowski}, \bibinfo{person}{John
  Richards}, {and} \bibinfo{person}{Kush~R Varshney}.}
  \bibinfo{year}{2020}\natexlab{}.
\newblock \showarticletitle{Experiences with improving the transparency of ai
  models and services}. In \bibinfo{booktitle}{\emph{Extended Abstracts of the
  2020 CHI Conference on Human Factors in Computing Systems}}.
  \bibinfo{pages}{1--8}.
\newblock


\bibitem[\protect\citeauthoryear{Hoffman, Mueller, Klein, and Litman}{Hoffman
  et~al\mbox{.}}{2018}]%
        {hoffman2018metrics}
\bibfield{author}{\bibinfo{person}{Robert~R Hoffman}, \bibinfo{person}{Shane~T
  Mueller}, \bibinfo{person}{Gary Klein}, {and} \bibinfo{person}{Jordan
  Litman}.} \bibinfo{year}{2018}\natexlab{}.
\newblock \showarticletitle{Metrics for explainable AI: Challenges and
  prospects}.
\newblock \bibinfo{journal}{\emph{arXiv preprint arXiv:1812.04608}}
  (\bibinfo{year}{2018}).
\newblock


\bibitem[\protect\citeauthoryear{Holland, Hosny, Newman, Joseph, and
  Chmielinski}{Holland et~al\mbox{.}}{2018}]%
        {holland2018dataset}
\bibfield{author}{\bibinfo{person}{Sarah Holland}, \bibinfo{person}{Ahmed
  Hosny}, \bibinfo{person}{Sarah Newman}, \bibinfo{person}{Joshua Joseph},
  {and} \bibinfo{person}{Kasia Chmielinski}.} \bibinfo{year}{2018}\natexlab{}.
\newblock \showarticletitle{The dataset nutrition label: A framework to drive
  higher data quality standards}.
\newblock \bibinfo{journal}{\emph{arXiv preprint arXiv:1805.03677}}
  (\bibinfo{year}{2018}).
\newblock


\bibitem[\protect\citeauthoryear{H{\"o}{\"o}k}{H{\"o}{\"o}k}{2000}]%
        {hook2000steps}
\bibfield{author}{\bibinfo{person}{Kristina H{\"o}{\"o}k}.}
  \bibinfo{year}{2000}\natexlab{}.
\newblock \showarticletitle{Steps to take before intelligent user interfaces
  become real}.
\newblock \bibinfo{journal}{\emph{Interacting with computers}}
  \bibinfo{volume}{12}, \bibinfo{number}{4} (\bibinfo{year}{2000}),
  \bibinfo{pages}{409--426}.
\newblock


\bibitem[\protect\citeauthoryear{Hummer, Muthusamy, Rausch, Dube, El~Maghraoui,
  Murthi, and Oum}{Hummer et~al\mbox{.}}{2019}]%
        {hummer2019modelops}
\bibfield{author}{\bibinfo{person}{Waldemar Hummer}, \bibinfo{person}{Vinod
  Muthusamy}, \bibinfo{person}{Thomas Rausch}, \bibinfo{person}{Parijat Dube},
  \bibinfo{person}{Kaoutar El~Maghraoui}, \bibinfo{person}{Anupama Murthi},
  {and} \bibinfo{person}{Punleuk Oum}.} \bibinfo{year}{2019}\natexlab{}.
\newblock \showarticletitle{Modelops: Cloud-based lifecycle management for
  reliable and trusted ai}. In \bibinfo{booktitle}{\emph{2019 IEEE
  International Conference on Cloud Engineering (IC2E)}}. IEEE,
  \bibinfo{pages}{113--120}.
\newblock


\bibitem[\protect\citeauthoryear{Ibitoye, Abou-Khamis, Matrawy, and
  Shafiq}{Ibitoye et~al\mbox{.}}{2019}]%
        {ibitoye2019threat}
\bibfield{author}{\bibinfo{person}{Olakunle Ibitoye}, \bibinfo{person}{Rana
  Abou-Khamis}, \bibinfo{person}{Ashraf Matrawy}, {and}
  \bibinfo{person}{M~Omair Shafiq}.} \bibinfo{year}{2019}\natexlab{}.
\newblock \showarticletitle{The Threat of Adversarial Attacks on Machine
  Learning in Network Security--A Survey}.
\newblock \bibinfo{journal}{\emph{arXiv preprint arXiv:1911.02621}}
  (\bibinfo{year}{2019}).
\newblock


\bibitem[\protect\citeauthoryear{Idowu, Str{\"u}ber, and Berger}{Idowu
  et~al\mbox{.}}{2021}]%
        {idowu2021asset}
\bibfield{author}{\bibinfo{person}{Samuel Idowu}, \bibinfo{person}{Daniel
  Str{\"u}ber}, {and} \bibinfo{person}{Thorsten Berger}.}
  \bibinfo{year}{2021}\natexlab{}.
\newblock \showarticletitle{Asset Management in Machine Learning: A Survey}.
\newblock \bibinfo{journal}{\emph{arXiv preprint arXiv:2102.06919}}
  (\bibinfo{year}{2021}).
\newblock


\bibitem[\protect\citeauthoryear{Iyer, Li, Li, Lewis, Sundar, and Sycara}{Iyer
  et~al\mbox{.}}{2018}]%
        {iyer2018transparency}
\bibfield{author}{\bibinfo{person}{Rahul Iyer}, \bibinfo{person}{Yuezhang Li},
  \bibinfo{person}{Huao Li}, \bibinfo{person}{Michael Lewis},
  \bibinfo{person}{Ramitha Sundar}, {and} \bibinfo{person}{Katia Sycara}.}
  \bibinfo{year}{2018}\natexlab{}.
\newblock \showarticletitle{Transparency and explanation in deep reinforcement
  learning neural networks}. In \bibinfo{booktitle}{\emph{Proceedings of the
  2018 AAAI/ACM Conference on AI, Ethics, and Society}}.
  \bibinfo{pages}{144--150}.
\newblock


\bibitem[\protect\citeauthoryear{Ji, Lipton, and Elkan}{Ji
  et~al\mbox{.}}{2014}]%
        {ji2014differential}
\bibfield{author}{\bibinfo{person}{Zhanglong Ji}, \bibinfo{person}{Zachary~C
  Lipton}, {and} \bibinfo{person}{Charles Elkan}.}
  \bibinfo{year}{2014}\natexlab{}.
\newblock \showarticletitle{Differential privacy and machine learning: a survey
  and review}.
\newblock \bibinfo{journal}{\emph{arXiv preprint arXiv:1412.7584}}
  (\bibinfo{year}{2014}).
\newblock


\bibitem[\protect\citeauthoryear{Lipton}{Lipton}{2018}]%
        {lipton2018mythos}
\bibfield{author}{\bibinfo{person}{Zachary~C Lipton}.}
  \bibinfo{year}{2018}\natexlab{}.
\newblock \showarticletitle{The Mythos of Model Interpretability: In machine
  learning, the concept of interpretability is both important and slippery.}
\newblock \bibinfo{journal}{\emph{Queue}} \bibinfo{volume}{16},
  \bibinfo{number}{3} (\bibinfo{year}{2018}), \bibinfo{pages}{31--57}.
\newblock


\bibitem[\protect\citeauthoryear{Liu, Li, Zhao, Cai, Yu, and Leung}{Liu
  et~al\mbox{.}}{2018}]%
        {liu2018survey}
\bibfield{author}{\bibinfo{person}{Qiang Liu}, \bibinfo{person}{Pan Li},
  \bibinfo{person}{Wentao Zhao}, \bibinfo{person}{Wei Cai},
  \bibinfo{person}{Shui Yu}, {and} \bibinfo{person}{Victor~CM Leung}.}
  \bibinfo{year}{2018}\natexlab{}.
\newblock \showarticletitle{A survey on security threats and defensive
  techniques of machine learning: A data driven view}.
\newblock \bibinfo{journal}{\emph{IEEE access}}  \bibinfo{volume}{6}
  (\bibinfo{year}{2018}), \bibinfo{pages}{12103--12117}.
\newblock


\bibitem[\protect\citeauthoryear{Mehrabi, Morstatter, Saxena, Lerman, and
  Galstyan}{Mehrabi et~al\mbox{.}}{2019}]%
        {mehrabi2019survey}
\bibfield{author}{\bibinfo{person}{Ninareh Mehrabi}, \bibinfo{person}{Fred
  Morstatter}, \bibinfo{person}{Nripsuta Saxena}, \bibinfo{person}{Kristina
  Lerman}, {and} \bibinfo{person}{Aram Galstyan}.}
  \bibinfo{year}{2019}\natexlab{}.
\newblock \showarticletitle{A survey on bias and fairness in machine learning}.
\newblock \bibinfo{journal}{\emph{arXiv preprint arXiv:1908.09635}}
  (\bibinfo{year}{2019}).
\newblock


\bibitem[\protect\citeauthoryear{M{\"o}ller and Hansson}{M{\"o}ller and
  Hansson}{2008}]%
        {moller2008principles}
\bibfield{author}{\bibinfo{person}{Niklas M{\"o}ller} {and}
  \bibinfo{person}{Sven~Ove Hansson}.} \bibinfo{year}{2008}\natexlab{}.
\newblock \showarticletitle{Principles of engineering safety: Risk and
  uncertainty reduction}.
\newblock \bibinfo{journal}{\emph{Reliability Engineering \& System Safety}}
  \bibinfo{volume}{93}, \bibinfo{number}{6} (\bibinfo{year}{2008}),
  \bibinfo{pages}{798--805}.
\newblock


\bibitem[\protect\citeauthoryear{Morley, Floridi, Kinsey, and Elhalal}{Morley
  et~al\mbox{.}}{2019}]%
        {morley2019overview}
\bibfield{author}{\bibinfo{person}{Jessica Morley}, \bibinfo{person}{Luciano
  Floridi}, \bibinfo{person}{Libby Kinsey}, {and} \bibinfo{person}{Anat
  Elhalal}.} \bibinfo{year}{2019}\natexlab{}.
\newblock \showarticletitle{From what to how. An overview of AI ethics tools,
  methods and research to translate principles into practices}.
\newblock \bibinfo{journal}{\emph{arXiv preprint arXiv:1905.06876}}
  (\bibinfo{year}{2019}).
\newblock


\bibitem[\protect\citeauthoryear{Nicolae, Sinn, Tran, Buesser, Rawat, Wistuba,
  Zantedeschi, Baracaldo, Chen, Ludwig, et~al\mbox{.}}{Nicolae
  et~al\mbox{.}}{2018}]%
        {nicolae2018adversarial}
\bibfield{author}{\bibinfo{person}{Maria-Irina Nicolae},
  \bibinfo{person}{Mathieu Sinn}, \bibinfo{person}{Minh~Ngoc Tran},
  \bibinfo{person}{Beat Buesser}, \bibinfo{person}{Ambrish Rawat},
  \bibinfo{person}{Martin Wistuba}, \bibinfo{person}{Valentina Zantedeschi},
  \bibinfo{person}{Nathalie Baracaldo}, \bibinfo{person}{Bryant Chen},
  \bibinfo{person}{Heiko Ludwig}, {et~al\mbox{.}}}
  \bibinfo{year}{2018}\natexlab{}.
\newblock \showarticletitle{Adversarial Robustness Toolbox v1. 0.0}.
\newblock \bibinfo{journal}{\emph{arXiv preprint arXiv:1807.01069}}
  (\bibinfo{year}{2018}).
\newblock


\bibitem[\protect\citeauthoryear{Norvill, Cassanges, Shbair, Hilger, Cullen,
  and State}{Norvill et~al\mbox{.}}{2020}]%
        {norvill2020security}
\bibfield{author}{\bibinfo{person}{Robert Norvill}, \bibinfo{person}{Cyril
  Cassanges}, \bibinfo{person}{Wazen Shbair}, \bibinfo{person}{Jean Hilger},
  \bibinfo{person}{Andrea Cullen}, {and} \bibinfo{person}{Radu State}.}
  \bibinfo{year}{2020}\natexlab{}.
\newblock \showarticletitle{A security and privacy focused kyc data sharing
  platform}. In \bibinfo{booktitle}{\emph{Proceedings of the 2nd ACM
  International Symposium on Blockchain and Secure Critical Infrastructure}}.
  \bibinfo{pages}{151--160}.
\newblock


\bibitem[\protect\citeauthoryear{Nyholm and Smids}{Nyholm and Smids}{2016}]%
        {nyholm2016ethics}
\bibfield{author}{\bibinfo{person}{Sven Nyholm} {and} \bibinfo{person}{Jilles
  Smids}.} \bibinfo{year}{2016}\natexlab{}.
\newblock \showarticletitle{The ethics of accident-algorithms for self-driving
  cars: An applied trolley problem?}
\newblock \bibinfo{journal}{\emph{Ethical theory and moral practice}}
  \bibinfo{volume}{19}, \bibinfo{number}{5} (\bibinfo{year}{2016}),
  \bibinfo{pages}{1275--1289}.
\newblock


\bibitem[\protect\citeauthoryear{Sanders and Saxe}{Sanders and Saxe}{2017}]%
        {sanders2017garbage}
\bibfield{author}{\bibinfo{person}{Hillary Sanders} {and}
  \bibinfo{person}{Joshua Saxe}.} \bibinfo{year}{2017}\natexlab{}.
\newblock \showarticletitle{Garbage in, garbage out: How purport-edly great ML
  models can be screwed up by bad data}.
\newblock \bibinfo{journal}{\emph{Proceedings of Blackhat 2017}}
  (\bibinfo{year}{2017}).
\newblock


\bibitem[\protect\citeauthoryear{Saria and Subbaswamy}{Saria and
  Subbaswamy}{2019}]%
        {saria2019tutorial}
\bibfield{author}{\bibinfo{person}{Suchi Saria} {and} \bibinfo{person}{Adarsh
  Subbaswamy}.} \bibinfo{year}{2019}\natexlab{}.
\newblock \showarticletitle{Tutorial: safe and reliable machine learning}.
\newblock \bibinfo{journal}{\emph{arXiv preprint arXiv:1904.07204}}
  (\bibinfo{year}{2019}).
\newblock


\bibitem[\protect\citeauthoryear{Schelter, Boese, Kirschnick, Klein, and
  Seufert}{Schelter et~al\mbox{.}}{2017}]%
        {schelter2017automatically}
\bibfield{author}{\bibinfo{person}{Sebastian Schelter},
  \bibinfo{person}{Joos-Hendrik Boese}, \bibinfo{person}{Johannes Kirschnick},
  \bibinfo{person}{Thoralf Klein}, {and} \bibinfo{person}{Stephan Seufert}.}
  \bibinfo{year}{2017}\natexlab{}.
\newblock \showarticletitle{Automatically tracking metadata and provenance of
  machine learning experiments}. In \bibinfo{booktitle}{\emph{Machine Learning
  Systems Workshop at NIPS}}. \bibinfo{pages}{27--29}.
\newblock


\bibitem[\protect\citeauthoryear{Siau and Wang}{Siau and Wang}{2018}]%
        {siau2018building}
\bibfield{author}{\bibinfo{person}{Keng Siau} {and} \bibinfo{person}{Weiyu
  Wang}.} \bibinfo{year}{2018}\natexlab{}.
\newblock \showarticletitle{Building trust in artificial intelligence, machine
  learning, and robotics}.
\newblock \bibinfo{journal}{\emph{Cutter Business Technology Journal}}
  \bibinfo{volume}{31}, \bibinfo{number}{2} (\bibinfo{year}{2018}),
  \bibinfo{pages}{47--53}.
\newblock


\bibitem[\protect\citeauthoryear{Sokol and Flach}{Sokol and Flach}{2020}]%
        {sokol2020explainability}
\bibfield{author}{\bibinfo{person}{Kacper Sokol} {and} \bibinfo{person}{Peter
  Flach}.} \bibinfo{year}{2020}\natexlab{}.
\newblock \showarticletitle{Explainability fact sheets: a framework for
  systematic assessment of explainable approaches}. In
  \bibinfo{booktitle}{\emph{Proceedings of the 2020 Conference on Fairness,
  Accountability, and Transparency}}. \bibinfo{pages}{56--67}.
\newblock


\bibitem[\protect\citeauthoryear{Souza, Azevedo, Louren{\c{c}}o, Soares,
  Thiago, Brand{\~a}o, Civitarese, Brazil, Moreno, Valduriez,
  et~al\mbox{.}}{Souza et~al\mbox{.}}{2019a}]%
        {souza2019provenance}
\bibfield{author}{\bibinfo{person}{Renan Souza}, \bibinfo{person}{Leonardo
  Azevedo}, \bibinfo{person}{V{\'\i}tor Louren{\c{c}}o}, \bibinfo{person}{Elton
  Soares}, \bibinfo{person}{Raphael Thiago}, \bibinfo{person}{Rafael
  Brand{\~a}o}, \bibinfo{person}{Daniel Civitarese}, \bibinfo{person}{Emilio
  Brazil}, \bibinfo{person}{Marcio Moreno}, \bibinfo{person}{Patrick
  Valduriez}, {et~al\mbox{.}}} \bibinfo{year}{2019}\natexlab{a}.
\newblock \showarticletitle{Provenance data in the machine learning lifecycle
  in computational science and engineering}. In \bibinfo{booktitle}{\emph{2019
  IEEE/ACM Workflows in Support of Large-Scale Science (WORKS)}}. IEEE,
  \bibinfo{pages}{1--10}.
\newblock


\bibitem[\protect\citeauthoryear{Souza, Azevedo, Thiago, Soares, Nery, Netto,
  Vital, Cerqueira, Valduriez, and Mattoso}{Souza et~al\mbox{.}}{2019b}]%
        {souza2019efficient}
\bibfield{author}{\bibinfo{person}{Renan Souza}, \bibinfo{person}{Leonardo
  Azevedo}, \bibinfo{person}{Raphael Thiago}, \bibinfo{person}{Elton Soares},
  \bibinfo{person}{Marcelo Nery}, \bibinfo{person}{Marco~AS Netto},
  \bibinfo{person}{Emilio Vital}, \bibinfo{person}{Renato Cerqueira},
  \bibinfo{person}{Patrick Valduriez}, {and} \bibinfo{person}{Marta Mattoso}.}
  \bibinfo{year}{2019}\natexlab{b}.
\newblock \showarticletitle{Efficient runtime capture of multiworkflow data
  using provenance}. In \bibinfo{booktitle}{\emph{2019 15th International
  Conference on eScience (eScience)}}. IEEE, \bibinfo{pages}{359--368}.
\newblock


\bibitem[\protect\citeauthoryear{Varshney and Alemzadeh}{Varshney and
  Alemzadeh}{2017}]%
        {varshney2017safety}
\bibfield{author}{\bibinfo{person}{Kush~R Varshney} {and} \bibinfo{person}{Homa
  Alemzadeh}.} \bibinfo{year}{2017}\natexlab{}.
\newblock \showarticletitle{On the safety of machine learning: Cyber-physical
  systems, decision sciences, and data products}.
\newblock \bibinfo{journal}{\emph{Big data}} \bibinfo{volume}{5},
  \bibinfo{number}{3} (\bibinfo{year}{2017}), \bibinfo{pages}{246--255}.
\newblock


\bibitem[\protect\citeauthoryear{Villaronga, Kieseberg, and Li}{Villaronga
  et~al\mbox{.}}{2018}]%
        {villaronga2018humans}
\bibfield{author}{\bibinfo{person}{Eduard~Fosch Villaronga},
  \bibinfo{person}{Peter Kieseberg}, {and} \bibinfo{person}{Tiffany Li}.}
  \bibinfo{year}{2018}\natexlab{}.
\newblock \showarticletitle{Humans forget, machines remember: Artificial
  intelligence and the right to be forgotten}.
\newblock \bibinfo{journal}{\emph{Computer Law \& Security Review}}
  \bibinfo{volume}{34}, \bibinfo{number}{2} (\bibinfo{year}{2018}),
  \bibinfo{pages}{304--313}.
\newblock


\bibitem[\protect\citeauthoryear{Voigt and Bussche}{Voigt and Bussche}{2017}]%
        {GDPR}
\bibfield{author}{\bibinfo{person}{Paul Voigt} {and} \bibinfo{person}{Axel
  von~dem Bussche}.} \bibinfo{year}{2017}\natexlab{}.
\newblock \bibinfo{booktitle}{\emph{The EU General Data Protection Regulation
  (GDPR): A Practical Guide} (\bibinfo{edition}{1st} ed.)}.
\newblock \bibinfo{publisher}{Springer Publishing Company, Incorporated}.
\newblock
\showISBNx{3319579584}


\bibitem[\protect\citeauthoryear{Yang, Steinfeld, Ros{\'e}, and Zimmerman}{Yang
  et~al\mbox{.}}{2020}]%
        {yang2020re}
\bibfield{author}{\bibinfo{person}{Qian Yang}, \bibinfo{person}{Aaron
  Steinfeld}, \bibinfo{person}{Carolyn Ros{\'e}}, {and} \bibinfo{person}{John
  Zimmerman}.} \bibinfo{year}{2020}\natexlab{}.
\newblock \showarticletitle{Re-examining whether, why, and how human-ai
  interaction is uniquely difficult to design}. In
  \bibinfo{booktitle}{\emph{Proceedings of the 2020 chi conference on human
  factors in computing systems}}. \bibinfo{pages}{1--13}.
\newblock


\bibitem[\protect\citeauthoryear{Yanisky-Ravid and Hallisey}{Yanisky-Ravid and
  Hallisey}{2019}]%
        {yanisky2019equality}
\bibfield{author}{\bibinfo{person}{Shlomit Yanisky-Ravid} {and}
  \bibinfo{person}{Sean~K Hallisey}.} \bibinfo{year}{2019}\natexlab{}.
\newblock \showarticletitle{Equality and Privacy by Design: A New Model of
  Artificial Intelligence Data Transparency Via Auditing, Certification, and
  Safe Harbor Regimes}.
\newblock \bibinfo{journal}{\emph{Fordham Urb. LJ}}  \bibinfo{volume}{46}
  (\bibinfo{year}{2019}), \bibinfo{pages}{428}.
\newblock


\bibitem[\protect\citeauthoryear{Zaharia, Chen, Davidson, Ghodsi, Hong,
  Konwinski, Murching, Nykodym, Ogilvie, Parkhe, et~al\mbox{.}}{Zaharia
  et~al\mbox{.}}{2018}]%
        {zaharia2018accelerating}
\bibfield{author}{\bibinfo{person}{Matei Zaharia}, \bibinfo{person}{Andrew
  Chen}, \bibinfo{person}{Aaron Davidson}, \bibinfo{person}{Ali Ghodsi},
  \bibinfo{person}{Sue~Ann Hong}, \bibinfo{person}{Andy Konwinski},
  \bibinfo{person}{Siddharth Murching}, \bibinfo{person}{Tomas Nykodym},
  \bibinfo{person}{Paul Ogilvie}, \bibinfo{person}{Mani Parkhe},
  {et~al\mbox{.}}} \bibinfo{year}{2018}\natexlab{}.
\newblock \showarticletitle{Accelerating the Machine Learning Lifecycle with
  MLflow.}
\newblock \bibinfo{journal}{\emph{IEEE Data Eng. Bull.}} \bibinfo{volume}{41},
  \bibinfo{number}{4} (\bibinfo{year}{2018}), \bibinfo{pages}{39--45}.
\newblock


\bibitem[\protect\citeauthoryear{Zhang, Sparks, and Franklin}{Zhang
  et~al\mbox{.}}{2017}]%
        {zhang2017diagnosing}
\bibfield{author}{\bibinfo{person}{Zhao Zhang}, \bibinfo{person}{Evan~R
  Sparks}, {and} \bibinfo{person}{Michael~J Franklin}.}
  \bibinfo{year}{2017}\natexlab{}.
\newblock \showarticletitle{Diagnosing machine learning pipelines with
  fine-grained lineage}. In \bibinfo{booktitle}{\emph{Proceedings of the 26th
  International Symposium on High-Performance Parallel and Distributed
  Computing}}. \bibinfo{pages}{143--153}.
\newblock


\end{thebibliography}

\end{document}